\documentclass{llncs}
\usepackage{llncsdoc} 
\usepackage{graphicx}

\begin{document}


\title{Machine Learning and Social Robotics for Detecting Early Signs of Dementia}

\author{Patrik Jonell$^1$, Joseph Mendelson$^1$, Thomas Storskog$^2$, G{\"o}ran Hagman$^2$, Per {\"O}stberg$^3$,  Iolanda Leite$^4$, Taras Kucherenko$^4$, Olga Mikheeva$^4$, Ulrika Akenine$^2$, Vesna Jelic$^2$, Alina Solomon$^2$, Jonas Beskow$^1$, Joakim Gustafson$^1$, Miia Kivipelto$^2$, Hedvig Kjellstr{\"o}m$^4$}

\institute{$^1$Dept.~Speech, Music and Hearing, KTH Royal Institute of Technology, Sweden\\
$^2$Dept.~Neurobiology, Care Sciences and Society, Karolinska Institute, Sweden\\
$^3$Dept.~Clinical Science, Intervention and Technology, Karolinska Institute, Sweden\\
$^4$Dept.~Robotics, Perception and Learning, KTH Royal Institute of Technology, Sweden\\~~~\\
{\tt\small \{pjjonell,josephme,iolanda,tarask,olgamik,beskow,jkgu,hedvig\}@kth.se, \{thomas.storskog,goran.hagman,per.ostberg,ulrika.akenine, vesna.jelic,alina.solomon,miia.kivipelto\}@ki.se}}

\maketitle

\begin{abstract}
This paper presents the EACare project, an ambitious multi-disciplinary collaboration with the aim to develop an embodied system, capable of carrying out neuropsychological tests to detect early signs of dementia, e.g., due to Alzheimer's disease. The system will use methods from Machine Learning and Social Robotics, and be trained with examples of recorded clinician-patient interactions. The interaction will be developed using a participatory design approach. We describe the scope and method of the project, and report on a first Wizard of Oz prototype. 
\end{abstract}

\section{Introduction}
 
With an increasing number of elderly in the population, dementia is expected to increase. Dementia disorders and related illnesses such as depression in the aging population are devastating for the quality of life of the afflicted individuals, their families, and caregivers. Moreover, the increase in such disorders imply enormous costs for the health sectors of all developed countries. There is still no cure available. However, studies, e.g.~\cite{kivipelto2017}, show that it is possible to reduce the risk for dementia in later life through a number of preventative measures. Early diagnoses and prevention measures are thus key to counteract dementia.

Diagnoses of dementia are now made by teams of expert clinicians with standardized neuropsychological tests (Section \ref{sec:clinical}) as well as EEG and MRI examinations. A number of additional factors in the behavior are taken into regard apart from the actual test scores and measurements (Section \ref{sec:additional}). Using these methods, dementia in early stages are often missed, since the time intervals between the sessions usually are quite large, and symptoms often come and go over time. 
Once a diagnosis has been made, the caring medical expert typically prescribes a specific therapeutic intervention of the type of memory training. Training sessions are often scarce and require the presence of a medically trained person. 

The main goal of the multi-disciplinary research described in this position paper, is to develop an embodied agent (Section \ref{sec:furhat}) capable of interacting with elderly people in their home, assessing their mental abilities using both standardized test and analysis of their non-verbal behavior, in order to identify subjects at the first stages of dementia.

The contributions of such a system to the diagnostics and treatment of dementia would be twofold: 
\begin{enumerate}
\setlength{\itemsep}{0pt}
\item Firstly, since the proposed embodied system would enable tests to be carried out much more often, at the test subject's home, subtle and temporarily appearing symptoms might be detected that are missed in regular screenings at the clinic. Moreover, the computerized test procedure makes it possible to collect quantitative measurements of the subject's cognitive development over time, enabling a clinician overseeing the test process to base their diagnosis on rich statistical data instead of few, more qualitative, test results. All this will make it possible to introduce preventive therapy much earlier.
\item Secondly, after a mild cognitive impairment or early dementia diagnosis, the proposed system may also be used for patients to train regularly at home as a complement to sessions with an expert, greatly enhancing the efficiency of the therapy.
\end{enumerate}

The research efforts needed to achieve these goals span over such diverse research areas of Social Robotics, Geriatrics, and Machine Learning, and include:
\begin{enumerate}
\setlength{\itemsep}{0pt}
\item a user- and caregiver-friendly embodied agent that carries out established neuropsychological tests. This includes the development of an adaptive and sustainable dialogue system, with elements of gamification that encourages the subject to carry out the tests as effectively as possible (Section \ref{sec:automating_clinical}).
\item pioneering Machine Learning solutions to achieve diagnostics from the user's performance in the neuropsychological tests, as well as their non-verbal behavior during the test (Section \ref{sec:automating_additional}).
\item systematic evaluation of the developed systems in both large and focused social groups, using feedback from each evaluation to guide further scientific research and design.
\end{enumerate}

Research efforts in this direction have been made before, e.g.~\cite{rudzicz2015}. The novelty of the present project with respect to this work is that we propose a more advanced interaction and a more sophisticated embodiment that supports, among other things, mutual gaze and shared attention.

\section{Survey of related work}

Several efforts have been made to investigate the acceptance of social robots by the elderly \cite{heerink2010}, some of them with a particular focus on home environments \cite{Forlizzi2004,smarr2014}. For example, through a series of questionnaires and interviews to understand the attitudes and preferences of older adults for robot assistance with everyday tasks, Smarr et al. \cite{smarr2014} found that this user group is quite open for robots to perform a wide range of tasks in their homes. Surprisingly, for tasks such as chores or information management (e.g., reminders and monitoring their daily activity), older adults reported to prefer robot assistance over human assistance. 

One of the main motivations of developing assistive robots for the elderly is the fact that this technology can promote longer independent living \cite{Forlizzi2004}. However, in-home experiments are still scarce because of privacy concerns and limited robot autonomy. One of the few exceptions is the work by de Graaf et al. \cite{deGraaf2017}, who investigated reasons for technology abandonment in a study where 70 autonomous robots were deployed in people’s homes for about six months. Regardless of age (their participant pool ranged from 8 to 77 years), they found that the main reasons why people stopped using the robot were lack of enjoyment and perceived usefulness. These findings indicate that involving the target users from the early stages of development can be crucial for the success and acceptance of social robots.  

While most of the human-robot interaction studies with older adults have been conducted with neurotypical participants \cite{wada2007,Beer2011}, a few authors have included patients with dementia in their research. Sabanovic et al. \cite{sabanovic2013}, for instance, evaluated the effects of PARO, a seal-like robot, in group sensory therapy for older adults with different levels of dementia. The results of a seven-week study indicate that the robot's presence contributed to participants’ higher levels of engagement not only with the robot but also with the other people in the environment. Similar findings were obtained by Iacono and Marti \cite{iacono2016}, who compared the effects of the presence of a PARO robot and a stuffed toy during a group storytelling task. Furthermore, the authors found that while in the presence of the robot, participants' stories were much more articulated in terms of number of words, characters and narrative details. 

Perhaps most similar to our work both in terms of the interaction modality (language-based) and condition of the user group (older adults diagnosed with dementia), Rudzicz et al.~\cite{rudzicz2015} conducted a laboratory Wizard of Oz experiment to evaluate the challenges in speech recognition and dialogue between participants and a personal assistant robot while performing daily household tasks such as making tea. Their results suggest that autonomous language-based interactions in this setting can be challenging not only because of speech recognition errors but also because robots will often need to proactively employ conversational repair strategies during moments of confusion.









\section{The Furhat dialogue system}
\label{sec:furhat}

\begin{figure}[t]
\centerline{\includegraphics[width=0.7\linewidth]{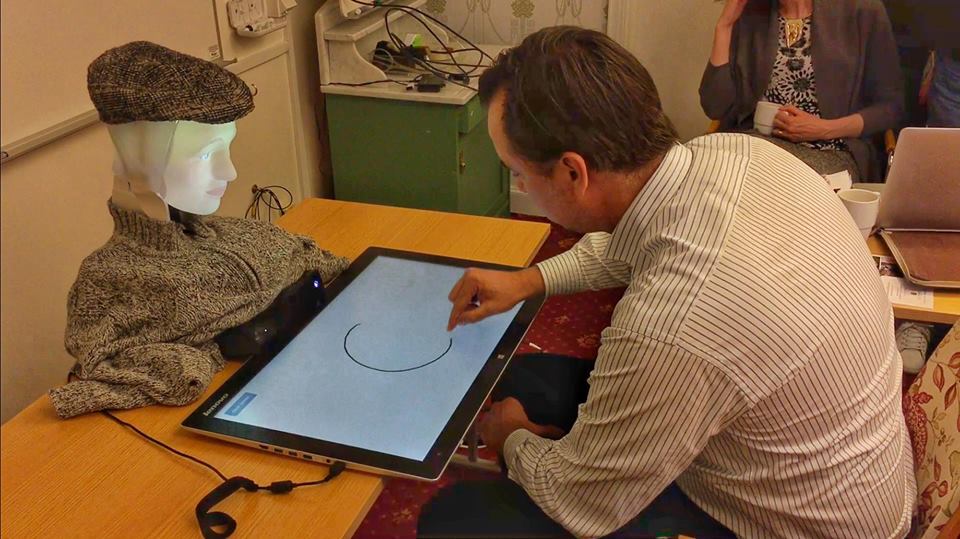}}
\caption{Furhat and one of the authors performing a clock test (Section \ref{sec:moca}).}
\label{fig:furhat}
\end{figure}

The project will make use of the robot platform Furhat \cite{almoubayed2013} (Figure \ref{fig:furhat}). Furhat is a unique social robotic head based on back-projected animation that has proven capable of exhibiting verbal and non-verbal cues in a manner that is in many ways superior to on-screen avatars. Furhat also offers advantages over mechatronic robot heads in terms of high facial expressivity, very low noise level and fast animation allowing e.g.~for highly accurate lip-synch. Since the face is projected on a mask, the robot’s visual appearance can be easily modified e.g.~to be more or less photo realistic or cartoonish, by changing the projection and/or the mask. This feature will be exploited in the participatory design study (Section \ref{sec:participatory}). 

The Furhat robot includes a multi-modal dialog system, implemented using the IrisTK framework \cite{skantze2012}. It uses statecharts as a powerful formalism for complex, reactive, event-driven spoken interaction management. It also includes a 3D situation model that makes it possible to handle situated interaction, including multiple users talking to the system and objects being discussed. A customizable Wizard of Oz mode as well as API:s for external control makes it straightforward to set up supervised experiments and data collection studies when a fully autonomous system can not be implemented.

\section{Clinical assessment of cognitive function}
\label{sec:clinical}

As described in the introduction, dementia is detected through clinical cognitive assessments. They are carried out under supervision of a clinician and evaluate attention, memory, language, visuospatial/perceptual functions, psycho- motor speed, and executive functions through a wide range of tests. Common and valid tests include learning and remembering a wordlist, e.g.~Rey Auditory Verbal Learning Test (RAVLT) \cite{woodard1999}, visuo-spatial and memory abilities, e.g.~Rey-Osterrieth Complex Figure Test (ROCF) \cite{mitrushina1990}, and test of mental and motor speed, e.g.~Wechsler Adult Intelligence Scale (WAIS) subtest Coding \cite{joy2004}.

Speech and language assessment includes, e.g., tests of picture naming \cite{kaplan1983}, word fluency \cite{tallberg2008}, and aspects of 
high-level language comprehension \cite{antonsson2016}.

\subsection{Montreal Cognitive Assessment test (MoCA)}
\label{sec:moca}

The Montreal Cognitive Assessment (MoCA) \cite{nasreddine2005} is a brief test that evaluates several of the cognitive domains mentioned above in a time-effective manner. Visuospatial and executive functions are here assessed using trail making, figure copying and clock drawing tasks. Language is assessed using object naming, sentence repetition, phonemic fluency and abstraction tasks. Attention is assessed using digit repetition, target detection and serial subtraction tasks. Memory is assessed using a delayed verbal recall task after initial learning trials. Temporal and spatial orientation is also assessed.

MoCA has proven robust in identifying subjects with mild cognitive impairment (MCI) and early Alzheimer’s disease (AD) and in distinguishing them from healthy controls, thereby becoming an important screening tool for clinicians all over the world \cite{julayanont2013}. 
Due to an increasing interest in using MoCA as a monitoring tool and the need of minimizing practice effects associated with repeated assessment, alternate forms of the test have also been developed 
\cite{costa2014}.

\section{Automating assessment of cognitive function}
\label{sec:automating_clinical}

We will in this project implement clinical neuropsychological assessments with the Furhat system (Section \ref{sec:furhat}). The Furhat agent will take the clinician's role, supervising and driving the test procedure. The assessments with Furhat will take place in the user's home and complement the regular assessments at the memory clinic. 

\subsection{Automating the Montreal Cognitive Assessment test (MoCA)}

We employ the Furhat system to implement the Montreal Cognitive Assessment test (MoCA). The intent is replicate the typical interaction between patient and clinician during the administration of the test, with the robot acting as the clinician. 

In order to rapidly get feedback from participants, a Wizard of Oz \cite{dahlback1993wizard} prototype has been developed where participants can interact with the embodied agent. The prototype consists of a Furhat robot and a touch-enabled tabletop computer (Figure \ref{fig:furhat}) connected to a remote wizarding interface. The system is set up to run through MoCA as described above. 

The prototype follows a scripted interaction, but also gives the wizard a large set of additional utterances in order to handle small deviations and make the interaction more life-like. These include typical conversational devices, such as affirmations, discourse markers, and the ability to add continuity or ``flow" to the interaction, such as ``ok, let's move on to the next section". The interaction is primarily centered around speech, which is realized by converting pre-set text prompts into speech output via a built in text-to-speech (TTS) system. However, some tasks require the participant to use the touch table in order to, for example, draw an image or recall the name of certain objects displayed on the screen. Besides providing the wizard with a set of utterances, the prototype also provides the wizard with the ability to perform certain facial expressions, head nods and sharing attention towards a given point. Using all of these modalities simultaneously allows the wizard to convey a highly engaging and believable embodiment of a clinician who can understand the user's actions and responses, and respond appropriately, just as a real clinician would.   

\begin{figure}[t]
\centerline{\includegraphics[width=0.9\linewidth]{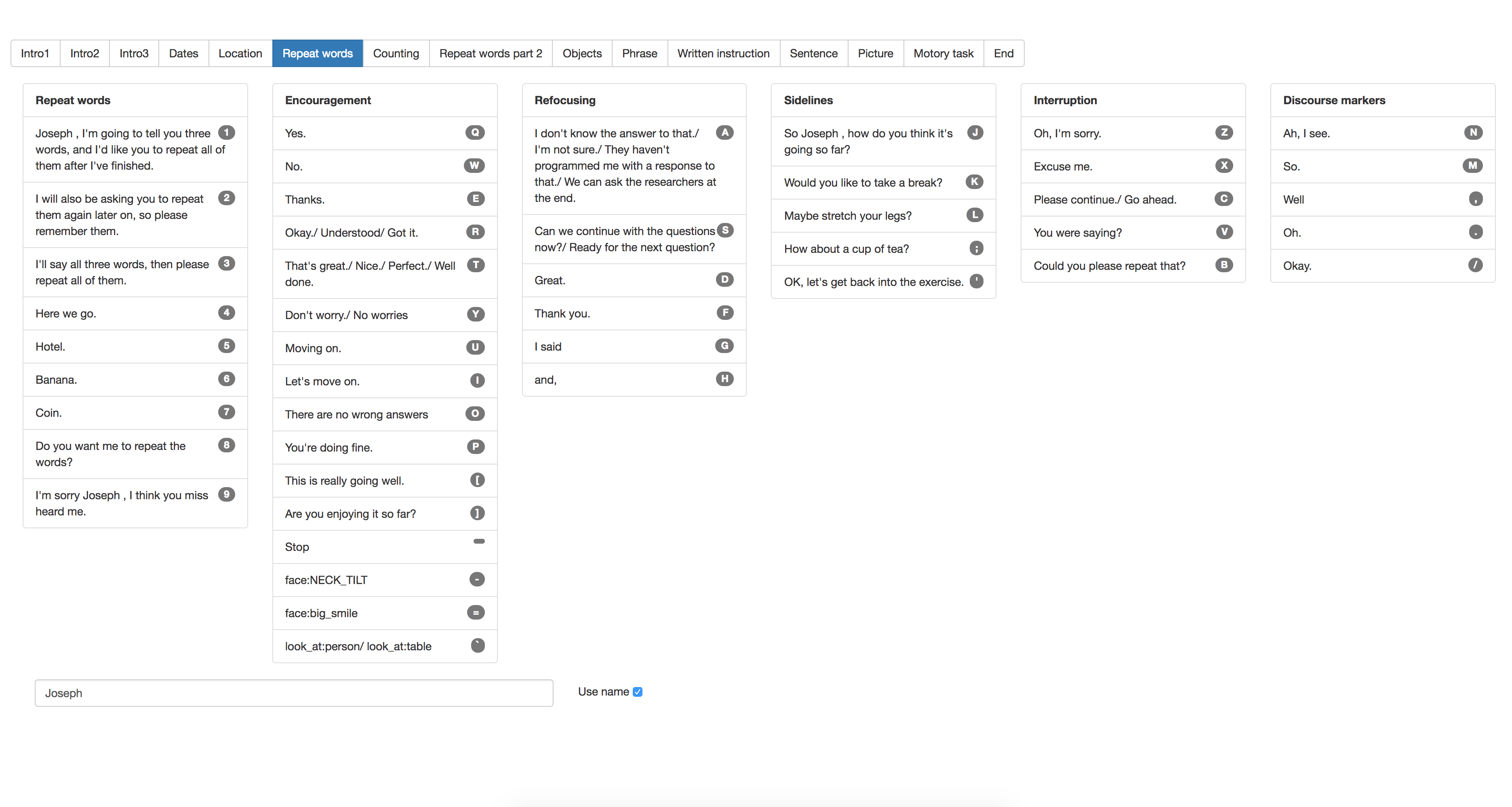}}
\caption{Wizard of Oz interface to the first Furhat system prototype.}
\label{fig:woz}
\end{figure}

The wizard interface consists of ``buttons" on a standard computer screen (Figure \ref{fig:woz}) that can be triggered from keys on its keyboard. These include the aforementioned pre-set text prompts which are realized as a TTS voice, as well as dedicated keys for facial expressions, head nods, and head movement to look at the touch screen. The wizard can hear participants via a microphone, and see them through a glass partition or via a webcam.

\subsection{Future Work: Participatory Design of the dialogue system}
\label{sec:participatory}

A series of workshops, each lasting 1-2 hours, will be conducted with around 10 participants.

These workshops will be based on the concept of Participatory Design, wherein the various stakeholders in the project will participate in the design of both the physical robot and the content of the proposed interactions. Stakeholders include potential users from the target demographic (volunteers and patients), clinicians with relevant expertise (KI), and the robot/interaction researchers (KTH). Specific activities during the workshops will include:

\begin{enumerate}
\item{introductory interaction with the robot system,}
\item{discussion/feedback of various aspects of system, e.g.~look, sound, content,}
\item{discussion of volunteers' and clinicians' preferences/expectations/concerns regarding system.}
\end{enumerate}

The process will be iterative: after each workshop session, the researchers will integrate ideas generated during the workshop into a new version of the system, which they will demonstrate at the subsequent session, which the stakeholders can then re-evaluate.

Interactions with the prototype will be recorded in order to improve the current system, but also in order to collect data used for building the automated system in the future (Section \ref{sec:automating_additional}). 

\section{Additional factors used in assessment of cognitive function}
\label{sec:additional}

During the clinical assessment, the clinician also make use of additional information in the diagnostic process.

Different neurodegenerative diseases often differ in socioemotional presentation; one example is mutual gaze \cite{sturm2011}. Studies have also shown changes in eye movements due to Alzheimer’s disease, with e.g.~alterations in gaze behavior \cite{molitor2015}.

Spoken interaction offers clues to cognitive functioning that are not usually measured or rated in clinical assessment. The temporal organization of speech, such as the incidence and duration of pauses, as well as the overall speaking rate, may signal word-finding problems and difficulties with discourse planning \cite{gayraud2011}. The prosodic organization, including pitch and loudness, may be abnormal in right-hemisphere brain lesions and in striatal loop dysfunction as in Parkinson’s disease and related cognitive disorders \cite{rektorova2016}.

\subsection{Future Work: Structured description of the diagnostic process}

To be able to incorporate such reasoning in the diagnostic process of the automatic dialogue system, we need to develop simplified and structured descriptions of what clinicians actually do during different kinds of assessment; pseudo-code-like scripts suitable for computer implementation.

These descriptions include both the verbal and non-verbal communication of the clinician during the cognitive assessment, but also the aspects of the patient's verbal and non-verbal communication relevant to assessment of cognitive function.

\section{Automating inference from additional factors}
\label{sec:automating_additional}

\begin{figure}[t]
\centerline{\includegraphics[width=0.7\linewidth]{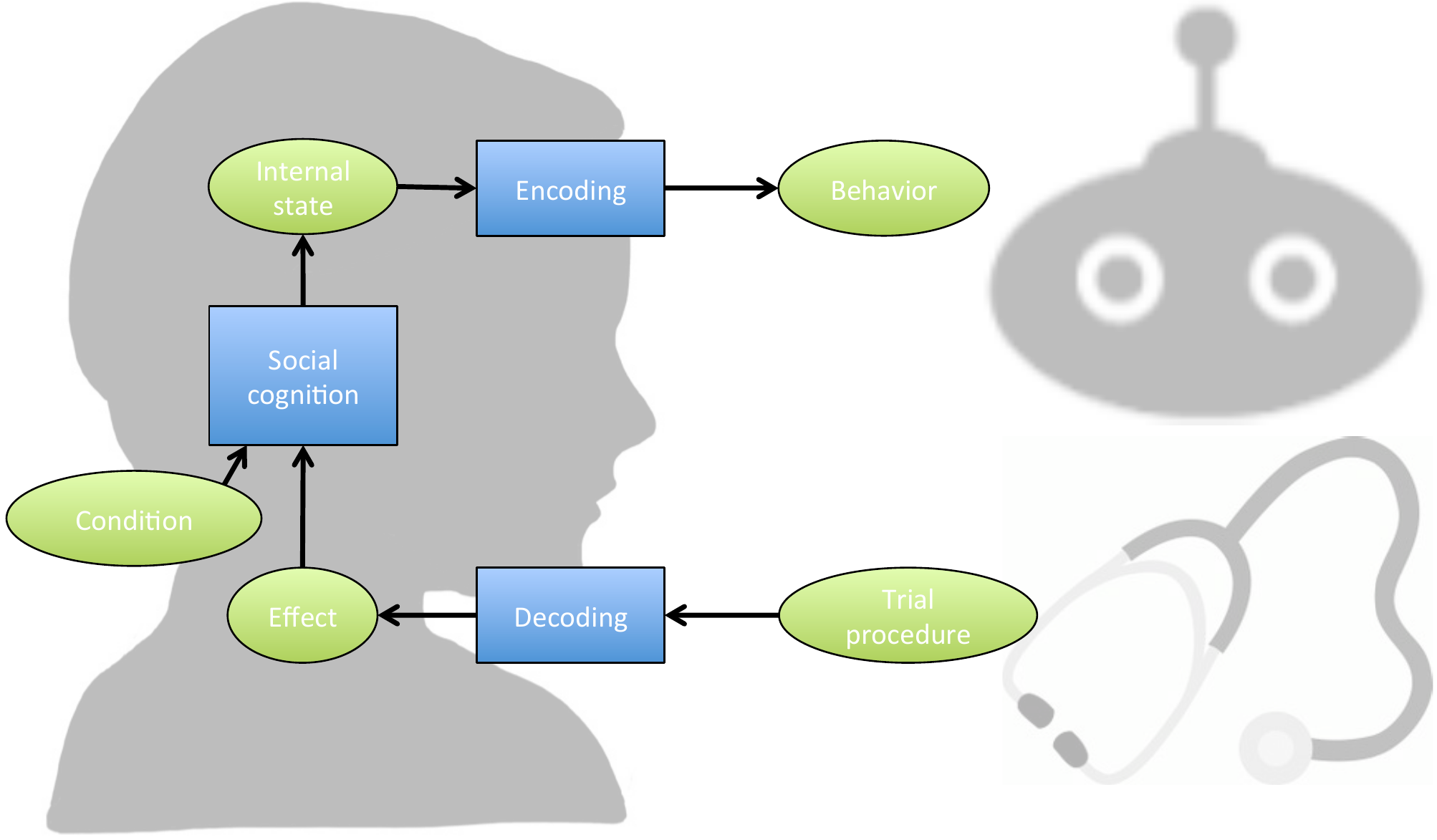}}
\caption{The diagnostics system will build upon a generative model of the patient's non-verbal behavior perception and production process.}
\label{fig:model}
\end{figure}

During the last decades, there has been a rapid development of methods for machine analysis of human spoken language. However, the information communicated between interacting humans is only to a small part verbal; humans transfer huge amounts of information through non-verbal cues such as face and body motion, gaze, and tone of voice, and these signals can be analyzed automatically to a certain degree \cite{burgoon2017}. 
As described in Section \ref{sec:additional}, dementia diagnostics relies heavily on such cues, and we aim to equip the system with the capability to take non-verbal cues into account in the diagnostics.

We will model the cognitive processes of non-verbal communication in the human brain (Figure \ref{fig:model}), on such a level that they explain the correlation between what the human perceives from the clinician's communication, and what the human in turn communicates. 
The underlying condition of an observed human can then be inferred from the recorded interaction with the clinician.

Data will be collected during user studies (Section \ref{sec:participatory}). Non-verbal signals of both clinician and evaluated human will be recorded using sensors such as Kinect human tracker, Tobii gaze detector, but also state-of-the-art techniques for extracting social signals from RGB-D video, e.g.~facial action units \cite{valstar2006}, and speech sound, e.g.~prosody, laughter and pauses \cite{gupta2013}.

\subsection{Future Work: Development of Machine Learning methodology}

The processes depicted in Figure \ref{fig:model} represent incredibly complex, non-smooth, and non-linear mappings and representations, which indicates that it will be suitable to use a deep neural network \cite{schmidhuber2014} approach.

Our initial studies \cite{butepage2017} show that it is beneficial to use generative, probabilistic deep models, in order to be able to incorporate prior information in a principled manner. Such information includes clinical expert knowledge and logic reasoning. 

Moreover, deep probabilistic approaches, such as~\cite{dai2016}, provide both more interpretability and lowers the needed amount of training data -- important aspects for a diagnostics method.

\section{Conclusions}

This position paper presented the EACare project, where we aim to develop a system with an embodied agent, that can carry out neuropsychological clinical tests and detect early signs of dementia from both the test results and from the user's non-verbal behavior.

This is intended as a stand-alone system, which the user brings home, and which interacts with the user in between regular screenings at the clinic. Two different ``spin-off products" could be 
\begin{enumerate}
\setlength{\itemsep}{0pt}
\item a system without embodiment or dialogue generation, which only serves as decision support to a clinician during a neuropsychological screening at the clinic, 
\item a system which passively monitors the user's communication behavior during daily activities, e.g.~as a back-end to Skype. 
\end{enumerate}
\vspace{-2mm}
\subsection*{Acknowledgements}
\vspace{-2mm}
The project described in this paper is funded by the Swedish Foundation for Strategic Research (SSF).

\vspace{-2mm}
\bibliographystyle{splncs03}
\bibliography{eacare} 

\end{document}